\documentclass{aa}  
\usepackage{graphicx}
\usepackage{txfonts}
\usepackage{natbib}
\usepackage{color}
\usepackage{hyperref}
\hypersetup{colorlinks=True,citecolor=blue,allcolors=blue}
\usepackage{tabularx}
\usepackage{tabu}
\usepackage{float}
\usepackage{multirow}
\pdfoutput=1 

\begin{document} 

   \title{Diffuse emission in microlensed quasars and its implications for accretion-disk physics}
   
\author{C. Fian\inst{1,2,3,4}, D. Chelouche\inst{2,3}, S. Kaspi\inst{4}}
\institute{Departamento de Astronom\'{i}a y Astrof\'{i}sica, Universidad de Valencia, E-46100 Burjassot, Valencia, Spain; \email{carina.fian@uv.es} \and Department of Physics, Faculty of Natural Sciences, University of Haifa,
Haifa 3498838, Israel \and Haifa Research Center for Theoretical Physics and Astrophysics, University of Haifa,
Haifa 3498838, Israel \and School of Physics and Astronomy and Wise Observatory, Raymond and Beverly Sackler Faculty of Exact Sciences, Tel-Aviv University, Tel-Aviv 6997801, Israel}
 

  \abstract
  {}
   {We investigate the discrepancy between the predicted size of accretion disks (ADs) in quasars and the observed sizes as deduced from gravitational microlensing studies. Specifically, we aim to understand whether the discrepancy is due to an inadequacy of current AD models or whether it can be accounted for by the contribution of diffuse broad-line region (BLR) emission to the observed continuum signal.}
   {We employed state-of-the-art emission models for quasars and high-resolution microlensing magnification maps and compared the attributes of their magnification-distribution functions to those obtained for pure Shakura-Sunyaev disk models. We tested the validity of our detailed model predictions by examining their agreement with published microlensing estimates of the half-light radius of the continuum-emitting region in a sample of lensed quasars.}
   {Our findings suggest that the steep disk temperature profiles found by microlensing studies are erroneous as the data are largely affected by the BLR, which does not obey a temperature-wavelength relation. We show with a sample of 12 lenses that the mere contribution of the BLR to the continuum signal is able to account for the deduced overestimation factors as well as the implied size-wavelength relation.}
   {Our study points to a likely solution to the AD size conundrum in lensed quasars, which is related to the interpretation of the observed signals rather than to disk physics. Our findings significantly weaken the tension between AD theory and observations, and suggest that microlensing can provide a new means to probe the hitherto poorly constrained diffuse BLR emission around accreting black holes.}


\titlerunning{Diffuse emission in microlensed quasars}
\authorrunning{Fian et al.} 
\maketitle
\section{Introduction} 

Accretion onto supermassive black holes (BHs) is tightly linked to their evolution over cosmic time. During the quasar phase, inflowing material forms a radiatively efficient accretion disk (AD), whose nature is still poorly understood. As accretion phenomena in quasars cannot be spatially resolved, indirect methods such as reverberation mapping (RM; \citealt{Blandford1982,Peterson1993}) and gravitational microlensing (\citealt{Wambsganss2006}) are required to probe them. 

Although thin disk models (hereafter SS73 disks) are an extremely useful prescription for modeling ADs and for estimating some of the fundamental properties of growing BHs, it has long been known that they, at least in their simplest form (\citealt{Shakura1973}), are inconsistent with quasar observations. Discrepancies between prevalent AD theory and observations are clearly manifested when considering the effective size of the continuum-emitting region in the UV-optical band, which is believed to be dominated by AD emission. While standard theory predicts that the effective wavelength-dependent size of the continuum-emitting region, $r(\lambda)$, increases with wavelength as $\lambda^{4/3}$, microlensing studies, which currently provide the only means to spatially resolve ADs in luminous high-redshift sources, often find a much flatter dependence across the UV-optical range (\citealt{Blackburne2011,Jimenez2014}). Furthermore, the implied AD sizes are much larger than predicted by theory, especially at rest-UV energies (\citealt{Motta2017,Cornachione2020a,Cornachione2020b,Cornachione2020c,Fian2016,Fian2018,Fian2021,Rojas2020}). This fact appears to be reinforced by RM studies of low-luminosity low-redshift sources, which employ a different set of assumptions and are prone to different biases (\citealt{Fausnaugh2018,Edelson2019,Chelouche2019,Fian2022,Fian2023}). It therefore appears to be a fairly robust conclusion that quasar continuum emission regions are larger than predicted by standard AD theory, and that their implied physics (e.g., the AD temperature profile) is markedly different from that expected from basic energetic arguments. This tension between quasar observations and theory has far-reaching implications for accretion physics in the general astrophysical context (\citealt{Hall2018,Li2019,Dexter2011,Gaskell2018,Gardner2017}) and in particular for explaining the census of supermassive BHs; considerable theoretical efforts are being made to remedy this (\citealt{Dexter2011,Hall2018,Li2019,Gaskell2018}).

Although it is commonly assumed that the UV-optical phenomenology of quasars is a good probe of AD physics, it has been theoretically argued (for more than 20 years) that the contribution of the broad-line region (BLR) to the continuum signal, albeit poorly constrained by observations, should be non-negligible, especially in low-luminosity sources (\citealt{Korista2001,Korista2019,Netzer2022}). Evidence has recently emerged for the non-negligible contribution of diffuse non-disk continuum emission to the time-varying signal of some low-luminosity, low-$z$ sources (\citealt{Cackett2018,Chelouche2019}), which is consistent with a BLR origin (\citealt{Korista2019,Chelouche2019,Netzer2020}). Nevertheless, it is as yet unclear whether the findings hold for all low-luminosity sources, or even for the quasar population as a whole, in which case the standard AD paradigm may be correct but our interpretation of the data would require a major revision (\citealt{Chelouche2019}). Alternatively, the findings may indicate that our understanding of both the AD and the BLR in quasars is highly incomplete, or even fundamentally flawed.

Motivated by this, we aim to explore the effect of diffuse BLR emission on size inferences from microlensing, which, unlike RM, is sensitive to the intrinsically varying and non-varying components of the nuclear signal. Doing so would test current models for the AD and the BLR at a substantially different part of the quasar phase space, for many more objects than are currently available through high-fidelity RM campaigns (\citealt{Kara2021,Kara2023}), and using a method whose systematic effects are very different from those inflicting RM. To this end, we employ state-of-the-art emission models for quasars (neglecting host emission) and high-resolution magnification maps, and compare attributes of their magnification-distribution functions to those obtained for pure SS73 disk models. 

The paper is organized as follows. In Section \ref{mlm} we describe the microlensing model and discuss its implementation in this work. Section \ref{qualitative} outlines a qualitative approach for modeling nuclear and diffuse quasar emission, and shows the emergence of biases in continuum-region size estimations due to the presence of the BLR. In Section \ref{quantitative} a physical  model for the BLR is presented, which is calibrated by spectroscopic and RM data, and its effect on the measured microlensing sizes is quantified and compared to published results for the lensed quasar sample of \citet{Blackburne2011}. Finally, in Section \ref{summary} we conclude by summarizing our key findings and discussing their broader implications.

\begin{figure}
\centering
\includegraphics[width=9cm]{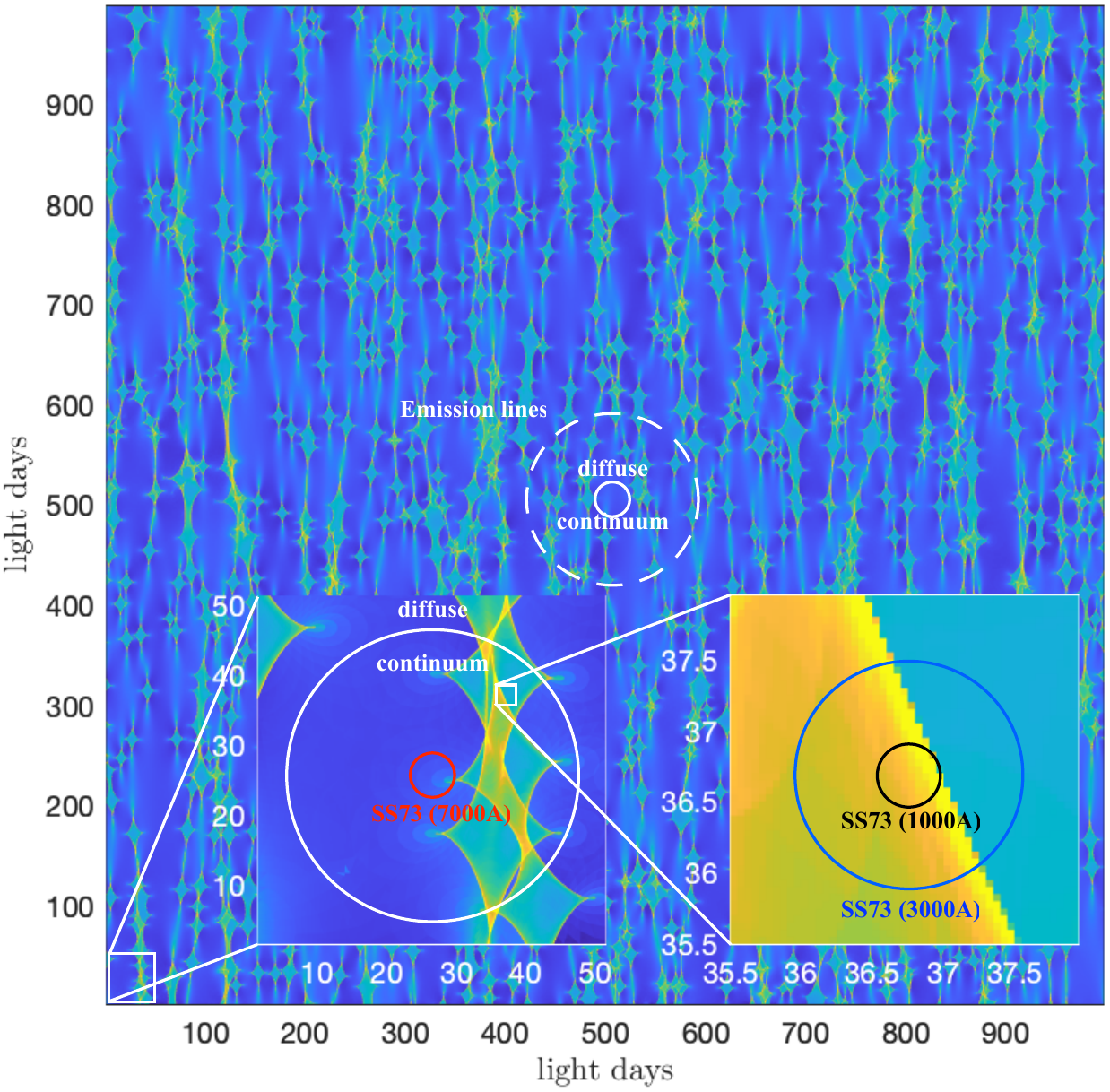}%
\caption{Billion-pixel magnification map showing a complex caustic pattern for the positive-parity image, with brighter colors indicating larger magnifications. The insets show successive blow-ups of the maps with relevant quasar emission regions overlaid.}
\label{ml_positive}
\end{figure}

\section{The microlensing model} \label{mlm}

Assessing the effect of diffuse BLR emission on microlensing signals involves several length scales defining the problem: the AD scale and the BLR scale (see Section \ref{qualitative} and \ref{quantitative}), as well as the relevant scales in the microlensing map associated with the Einstein radius, $r_E$, of the lensing stars and their field density (convergence, or optical depth for lensing), which we now outline. 

We simulate the effect of microlenses on the total macro-image flux using magnification maps calculated by the inverse polygon mapping (IPM) method (\citealt{Mediavilla2006,Mediavilla2011}). The IPM method is based on the properties of polygon transformation under the inverse mapping defined by the lens equation. The IPM achieves a remarkably high level of accuracy (relative error $\sim5\times10^{-4}$), accomplishing this in a significantly shorter computing time (less than 4\%) compared to the inverse ray shooting (IRS) method (\citealt{Wambsganss1999}).
The computational efficiency improvement of the IPM over the IRS can exceed two orders of magnitude when both methods are required to deliver the same level of accuracy.

The general characteristics of a magnification map are determined (for each quasar image) by the local convergence, $\kappa$, and the local shear, $\gamma$, which can be obtained by fitting a model to the coordinates of the macro images. The local convergence is proportional to the total surface mass density $\kappa=\kappa_c+\kappa_\star$, where $\kappa_c$ is the convergence caused by continuously distributed matter (e.g., dark matter) and $\kappa_\star$ can be attributed to stellar-mass point lenses. High convergence signifies a large mass concentration and, consequently, significant magnification. Conversely, lower convergence signifies a lower mass concentration, leading to reduced light bending and suppressed magnification. As such, convergence is pivotal in establishing the amplitude of the magnification map. The shear affects the shape of the magnification map, causing a distortion effect that manifests as a stretching along one axis, coupled with a compression along its perpendicular counterpart. Greater shear values result in highly extended features in the magnification map, whereas lower shear values induce smaller distortions, leading to features in the map that maintain a closer resemblance to their original forms.

We used the latest estimate for the fraction of the surface mass density in the form of stars ($\sim20\%$ near the Einstein radius of the lenses; see \citealt{Jimenez2015}). The maps are $20,000\times20,000$ pixels$^2$ large with sides of length $100r_E$ ($r_E\sim 3\times 10^{16}$\,cm; c.f. \citealt{Blackburne2011}) and a pixel size of 0.05\,light-days, which is smaller than the typical size of the UV AD (see Fig. \ref{ml_positive}). We then randomly distributed microlenses of mass $M_\star = 1\,M_\odot$ across the magnification patterns to create the convergence in stars $\kappa_\star$. Although the details of the microlensing pattern change with the shape of the initial mass function of the lens galaxy, the magnification probability distributions are insensitive to the choice of the initial mass function (\citealt{Lewis1995,Wyithe2001,Chan2021}). The source sizes can be scaled to a different stellar mass by noting that $r_E \propto M_\star^{1/2}$. The value at each point in the map is equal to the magnification of the background source at that point, relative to the average macro-image magnification. 

Characterizing the statistics of microlensing magnifications, we examined two typical cases: a positive-parity strongly lensed quasar image (minimum of the time-delay function) with $\kappa=\gamma=0.4$, and a theoretical average magnification of $\mu = 5$; and a negative-parity quasar image (saddle point) with $\kappa=\gamma=0.6$, and $\mu = -5$. The positive-parity simulation includes $\sim14,300$ stellar lenses, and the negative-parity simulation includes $\sim21,500$ stellar lenses. Microlensing-induced magnifications of our fiducial quasar are obtained by convolving its emission region (see below) with the resulting magnification map (see Fig. \ref{ml_positive}). In cases where the size of the emission is comparable to the size of the microlensing map and the derived statistic is questionable (this occurs when considering, for example, BLR emission from the most luminous quasars), calculations are reported using a lower resolution map with a pixel size of 0.25 light-days and a size of 400$r_E$ (not shown) while holding the other parameters fixed at their aforementioned values. 

\begin{figure}[t]
\includegraphics[width=8.9cm]{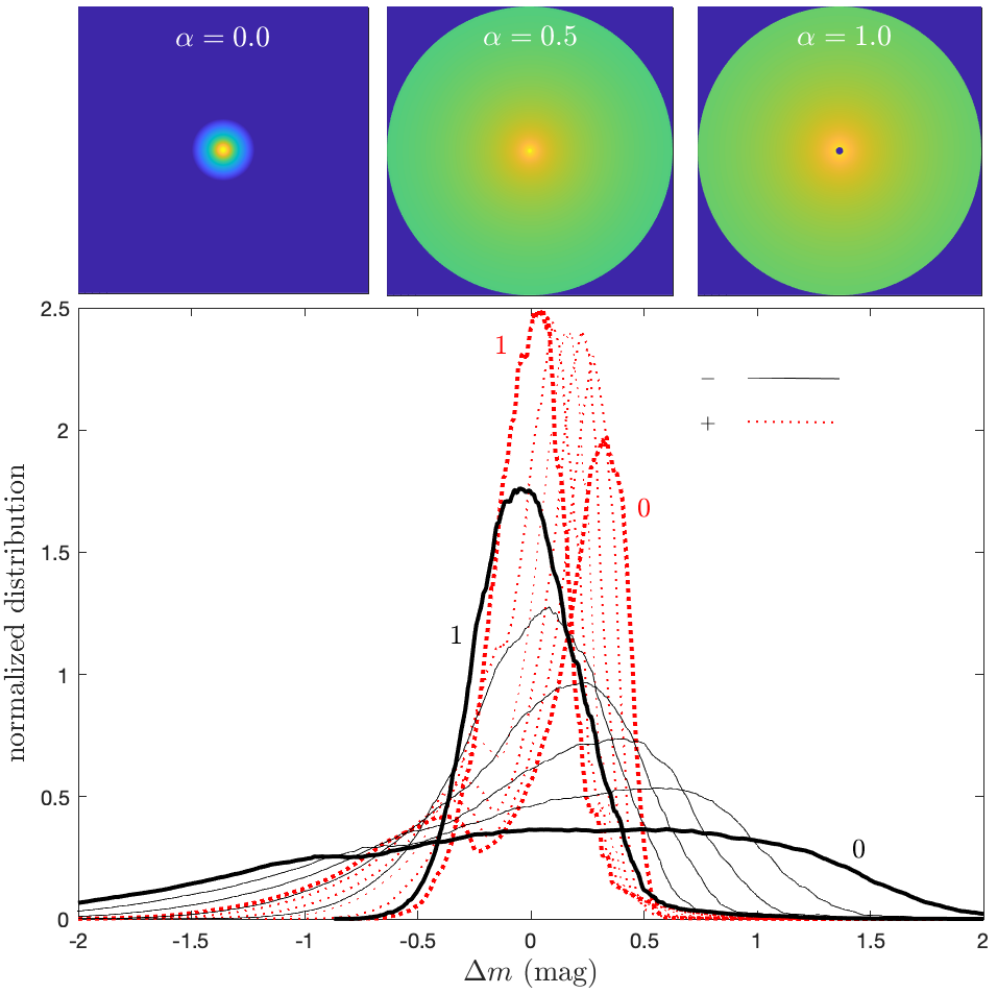}
\caption{Magnification distributions (relative to the mean magnification of the image) for several assumptions regarding the contribution of a diffuse emission component to the flux in some band, which is parameterized by $\alpha$ (see the main text). Bold distribution curves mark $\alpha=0.0$ and $\alpha=1.0$ cases for each image parity (see the legend), with the thin curves separated by $\Delta \alpha =0.2$ intervals. The top panels show the emissivity profiles in log-units for three models, with their $\alpha$-values denoted. Color-coding is arbitrarily but identically normalized in all panels. Note the reduced second moment of the magnification distribution function with increasing $\alpha$.}
\label{ml_PDF}
\end{figure}

Below we quantify the bias in the estimation of the AD size when diffuse (non-AD) emission is present but is not accounted for by the models that are employed to infer the disk size from the data. To this end we convolve models for quasar emission that include the AD and the BLR signals (henceforth SS73$+$BLR models; see below) with the computed magnification maps, and quantify the variance of the  magnification-distribution function (Fig. \ref{ml_PDF}), $\mathrm{Var_{SS73+BLR}}$. This is, essentially, a measure of the  observed microlensing-induced variability for a source that fully samples the microlensing map. While this may not be strictly true for particular sources observed over short times\footnote{A compact source could, for example, inhabit a particularly calm region of the microlensing map, leading to little microlensing induced variability, thereby behaving as a much larger source occupying a more typical region of the map.}, it is expected to be a more meaningful measure when compiling the results for a sample of lensed quasars, which better sample the landscape of their microlensing maps. This approach is also motivated by the \citet{Blackburne2011} sample used here, for which only single epoch measurements are available (see Section \ref{quantitative}). Ignoring prior knowledge of the model used to simulate the data, we next seek a pure SS73-disk model of half-light radius $r_{1/2}$, whose convolution with the same magnification map yields a variance $\mathrm{Var_{SS73}}(r_{1/2})$ that minimizes $\zeta(r_{1/2})=\vert \mathrm{Var_{SS73}}(r_{1/2})-\mathrm{Var_{SS73+BLR}} \vert / \mathrm{Var_{SS73+BLR}}$, thereby yielding the recovered $r_{1/2}$. This may then be contrasted with the values for the effective half-light radius and the AD scale used for the AD$+$BLR model (see below).

We note that in all cases explored here, the approach resulted in a well-defined minimum for $\zeta$, and was able to accurately recover the input value when a pure SS73 disk was used to model the source. We verified that our approach is also consistent with a maximum likelihood scheme that aims to find an optimal match between calculated magnification distribution functions, where a sizeable ($\pm 1$\,mag) uniform prior is used on the mean magnitude, and an extremum is sought \citep{Fian2021}. Our approach is further warranted by the current data quality for lensed quasars;  higher moments of the magnification distribution, which are not explored here, may be accessible to future surveys. 

\section{A qualitative emission model for quasars} \label{qualitative} 

To qualitatively assess the effect of diffuse non-AD emission on the recovered AD properties, we first consider a simplified approach wherein the normalized surface brightness (SB) over a narrow range of wavelength, $\Delta \lambda/\lambda \ll 1$, satisfies
\begin{equation}
I(R)=\frac{1-\alpha}{I_1}\left [ {\mathrm{exp}\left (R^{3/4} \right )-1} \right ]^{-1} + \frac{\alpha}{I_2} R^\beta  \chi_{[r_\mathrm{in},r_\mathrm{out}]}
\label{emm}
, \end{equation}
where the normalized radial coordinate, $R=r/r_s$, and $r_s$ is the wavelength-dependent disk scale. The first/second term in Eq. \ref{emm} describes the AD/BLR contribution to the signal, whose weight is parameterized by $\alpha \in [0,1]$, which may be wavelength-dependent (see below). Unless otherwise stated, we neglect the inner and outer disk boundaries. For an SS73 disk, the half-light radius of the disk, $r_{1/2}\simeq 2.44 r_s$ (c.f. $r_{1/2} \simeq 1.18r_s$ for a Gaussian disk), is set by the monochromatic rest optical luminosity, $L_\mathrm{opt}$, of the source up to an unknown inclination angle effect (a face on configuration is assumed throughout this work), and so long as the emission is radiated by the self-similar parts of the disk (see \citealt{Davis2011,Laor2011}). In this case, 
\begin{equation}
r_{1/2}\simeq 1.5 \left ( \frac{L_\mathrm{opt}}{10^{45}\,\mathrm{erg~s^{-1}}} \right )^{1/2} \left ( \frac{\lambda}{5100\mathrm{\AA}} \right )^{4/3}\,\mathrm{light\,days}.
\label{r1/2}
\end{equation}
Therefore, for a fiducial quasar with an optical luminosity, $L_\mathrm{opt,45}=L_\mathrm{opt}/10^{45}\,\mathrm{erg\,s^{-1}}=1$ \citep{Morgan2010,Blackburne2011}, $r_s=r_{s,3000}(\lambda/3000\mathrm{\AA})^{4/3}$ with $r_{s,3000}\simeq 0.3$\,light-days. The normalization factors in Eq. \ref{emm}, $I_1 \equiv \int_0^\infty \left [ {\mathrm{exp}\left (R^{3/4} \right )-1} \right ]^{-1} R\, dR$, $I_2 \equiv \int_0^\infty R^{\beta+1}  \chi_{[r_\mathrm{in},r_\mathrm{out}]}(R)\, dR$, where the step function $\chi_{[r_\mathrm{in},r_\mathrm{out}]}(R)=1$ if $r\in [r_\mathrm{in},r_\mathrm{out}]$ and is zero otherwise. The parameter $\beta$ sets the radial dependence of the diffuse SB component, and is taken here to be in the range $[-3,0]$, but is likely to be closer to $\beta \sim -2$ for most continuum emission components of interest (see \citealt{Korista2019}). Specifically, the SB depends on the combined effect of the local emissivity of the gas (as follows from its illumination by the central source), its geometrical configuration (e.g., the covering factor over the quasar sky), and radiative transfer effects along the sightline to the observer (absorption and scattering). In our formalism a three-dimensional emissivity profile, $\epsilon(r)\propto r^\gamma$ with a radial geometrical covering factor of $C(r)\propto r^\delta$ would translate to a radial dependence of the SB, which is $\propto r^\beta$ with $\beta=\gamma+\delta+1$. Therefore, a typical emissivity profile with $\gamma=-2$ (\citealt{Korista2019}), and $\delta=-1$ (\citealt{Baskin2014}), translates to $\beta=-2$. 

 For the diffuse BLR component, we assumed an inner radius $r_\mathrm{in}=70r_{s,3000}$ and an outer radius $r_\mathrm{out}=2\times 10^3r_{s,3000}$, which translates to a half-light radius of the BLR emission of $\sim 20$\,light-days for $\beta=-2$ and a half-light radius of $\sim 90$\,light-days for $\beta=0$. The latter values may be more appropriate for low-ionization emission lines (see \citealt{Korista2019}) and in particular the Balmer lines, whose emission may be collisionally suppressed at small radii (\citealt{Baskin2014}). Therefore, the model is qualitatively consistent with the BLR size-luminosity relation (\citealt{Bentz2013}).\\ 

\begin{figure}[t]
\centering
\includegraphics[width=8.9cm]{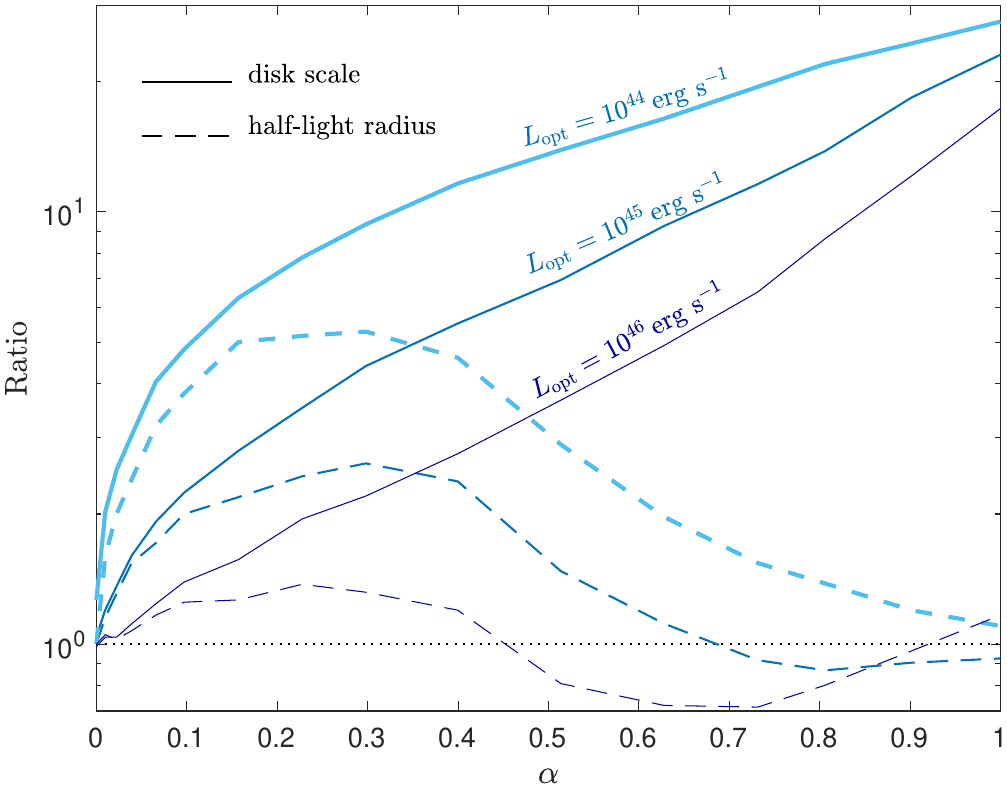}
\caption{Simulations quantifying the overestimation factors of the disk scale and the half-light radius as a function of the relative contribution of a diffuse BLR emission component to the total continuum signal, $\alpha$. The disk-scale overestimation factor is defined as the ratio of the value recovered via the magnification-distribution function that matches the input value, $R_s$, in our AD-BLR model. Results for the half-light radius are defined by the ratio of the value recovered for the best-matched pure AD model to that which characterizes the AD-BLR model.} 
\label{biases}
\end{figure}
\begin{figure}[t]
\centering
\includegraphics[width=8.9cm]{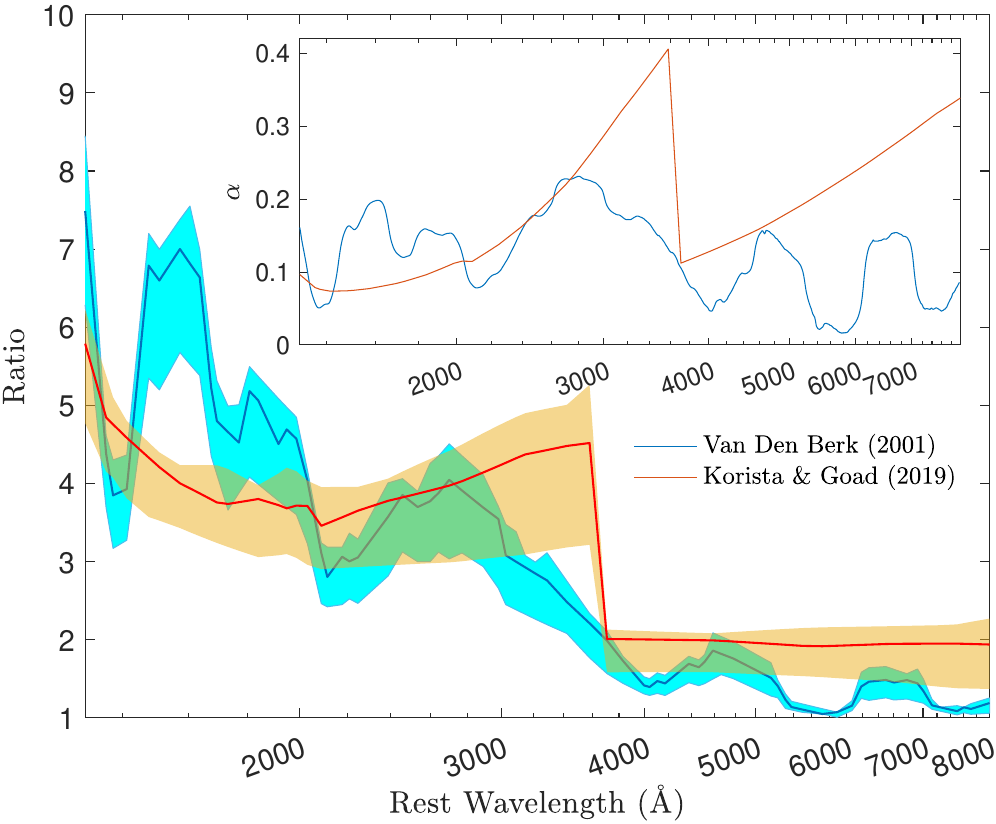}
\caption{Wavelength-dependent overestimation factors of disk sizes in the presence of a BLR component. The effect of the power-law index of the emissivity profile (Eq. \ref{emm}) is small within the range $[-3,0]$ (shaded regions). The inset shows the relative contribution of the BLR to the flux in the band, $\alpha$, according to two models: the observationally motivated phenomenological line emission model of \citet{VandenBerk2001} is shown in blue, and the more physical BLR model of \citet{Korista2019} in red.}
\label{kor_alpha}
\end{figure}

As expected, we find that the magnification distributions narrow down with rising $\alpha$ (Fig. \ref{ml_PDF}). The details of the magnification distributions are different for the positive and negative parity images of the quasar, and yet they lead to similar conclusions. Hence, only the negative parity results will be discussed below. Not unexpectedly, larger emitting regions average over more extended parts of the magnification map\footnote{A finite dispersion of the magnification distributions is also apparent in the limiting case of $\alpha=1$, thereby distinguishing the BLR model from highly extended sources, such as the host galaxy, whose contribution has not been included in the model.}, which suppress significant (de-)magnification events. Qualitatively, a best-match criterion between the magnification-distribution function that characterizes AD-BLR models and those of pure-disk models that are (erroneously) used to recover a disk size, would lead to overestimations. More quantitatively, the implied disk scale, $r_s$, in the presence of diffuse BLR emission can be several times larger than the true AD scale of the system for our fiducial quasar with an optical luminosity, $L_\mathrm{opt}=10^{45}\,\mathrm{erg~s^{-1}}$, even for modest values of $\alpha\sim 0.2$ (see Fig. \ref{biases}), which are motivated by photoionization calculations  (see \citealt{Korista2019} and the inset of our \mbox{Fig. \ref{kor_alpha}}). Interestingly, non-negligible biases also emerge when comparing the calculated half-light radius for the AD-BLR model to that implied by the best-matched pure-disk one, with overestimations factors of about two for typical values of $\alpha$. This result contrasts previous works, which often conjecture that the emissivity profile has little effect on half-light size estimations (see \citealt{Mortonson2005}).

The bias level in source-size estimations depends on the interplay between the size of quasar constituents and the Einstein radius for a fixed $\alpha$. To quantify this effect, we consider high-luminosity ($L_\mathrm{opt}=10^{46}\,\mathrm{erg~s^{-1}}$) and low-luminosity ($L_\mathrm{opt}=10^{44}\,\mathrm{erg~s^{-1}}$) quasars. We assume that the AD and BLR sizes scale as $L_\mathrm{opt}^{1/2}$ (see \citealt{Bentz2013}), and show the results in Fig. \ref{biases}. An overestimation of the disk size by factors of $\lesssim 10$ is obtained for low-luminosity sources. Size inferences for high-luminosity sources are less biased and are overestimated by $\sim 50\%$ for otherwise similar model parameterizations. Similar behavior with source luminosity is seen for the half-light size estimations' bias.

More generally, the source size and the relative contribution of the diffuse BLR component to the signal are wavelength-dependent for a given quasar, as is the bias. We demonstrate this for our fiducial quasar by considering two BLR models that are implemented using Eq. \ref{emm} while holding all other BLR properties fixed at their aforementioned values: a phenomenological model based on the quasar composite spectrum of  \citet{VandenBerk2001} and a physical model for the BLR emission that is based on photoionization calculations (\citealt{Korista2001,Korista2019}). In the former model, a broken power-law is fit to the data, and the residuals, normalized to the local power-law fit, are attributed to the BLR emission (\citealt{VandenBerk2001}) and hence set $\alpha(\lambda)$. A smoothed version of the model is shown in the inset of \mbox{Fig. \ref{kor_alpha}} to allow an easier comparison with photometric data. In the second model, an independent estimate for $\alpha(\lambda)$ is provided, which relies on the physical locally optimally emitting cloud model for the BLR (\citealt{Baldwin1995}), whose $\alpha(\lambda)$ predictions factor in many relevant processes, but do not include emission lines and line blends (\citealt{Korista2019}). In both of the models, $\alpha(\lambda)\gtrsim 0.1$ over much of the UV-optical range and may reach $\alpha\simeq 0.4$ near the radiative recombination-edge thresholds of the Balmer and Paschen series. These models are used in conjunction with Eq. \ref{emm} and the simulated magnification map to predict the biases shown in Fig. \ref{kor_alpha}. 

The two models for $\alpha(\lambda)$, despite being very different in construction, lead to qualitatively similar conclusions. The wavelength-dependent overestimation factor of the inferred disk size for our fiducial quasar is very substantial at rest-UV energies (Fig. \ref{kor_alpha}), where the disk emission is more compact. The bias is clearly affected by major spectral BLR features such as radiative-recombination edges and strong lines and line blends. Specifically, overestimation factors of $\gtrsim 4$ are expected at $\lesssim 4000$\AA\ where the small blue bump dominates (\citealt{Wills1985}). The contribution from the Rayleigh-scattering wings of Ly$\alpha$ (see the wavelength range below $\sim$1500\AA), may lead to significant biases according to these models. The Paschen edge at $\sim 8000$\AA\, leads to a smaller bias than the Balmer edge because the underlying disk scale at long wavelengths is larger. As demonstrated above, we expect sources whose luminosity is greater (smaller) than the fiducial value to be prone to smaller (larger) biases.

Varying the emissivity profile of the diffuse emission component (i.e., $\beta$) has a modest effect on the bias level for $\alpha<0.5$. For example, a $\beta=0$ model shifts the emission to larger radii, thereby increasing the bias by $\sim 10$\% relative to our fiducial case. Conversely, a steeper emissivity profile results in a somewhat reduced bias level. For narrowband data, where $\alpha>0.5$ values are encountered at the spectral locations of prominent emission lines, the implied half-light radius may be underestimated by a factor of a few for $\beta>-2$ (not shown).

\section{A quantitative assessment of reported AD sizes}  \label{quantitative}

Here we aim to check whether reported size estimations for the UV-optical disk in quasars, which largely disagree with AD models for these objects, could be reconciled with SS73 theory when considering the biases introduced by  diffuse BLR emission. A refined BLR model is presented below, followed by a comparison between AD size estimations for the \citet{Blackburne2011} sample and the simulations that include BLR emission. We emphasize that it is not our intention to solve the full observational problem for particular lenses, which is exacerbated by (unknown) intrinsic quasar variations with time-delays between the images, and is subject to uncertainties in the macro- and micro-lens models. 

\subsection{A realistic model for the BLR}  

We aim to incorporate a realistic model for the diffuse emission from the BLR, which includes all major emission lines, line-blends, and continuum emission, and is consistent with BLR phenomenology of major emission lines across a wide energy range. The BLR model considered here assumes radiation-pressure--confined (RPC) non-accelerating clouds \citep{Dopita2002}, which are ionization-bound, and for which emission and scattering from deep within the cold neutral regions of the clouds is neglected \citep{Baskin2014}. Specifically, a one-dimensional slab is illuminated from one side by the ionizing disk and corona radiation, thereby exerting radiation pressure force on the illuminated layers. Under hydrostatic equilibrium, pressure increases within the deeper cloud layers to counter-balance the radiation pressure force, thereby leading to high compression rates, and to the gas pressure in the optically thick cloud layers being comparable to the radiation pressure, as observed \citep{Baskin2014}. A salient feature of such models is a wide range of ionization levels characterizing individual clouds: highly ionized gas near the illuminated surface of the cloud to neutral gas deep in its dense and optically thick layers.

\begin{figure}[t]
\includegraphics[width=8.9cm]{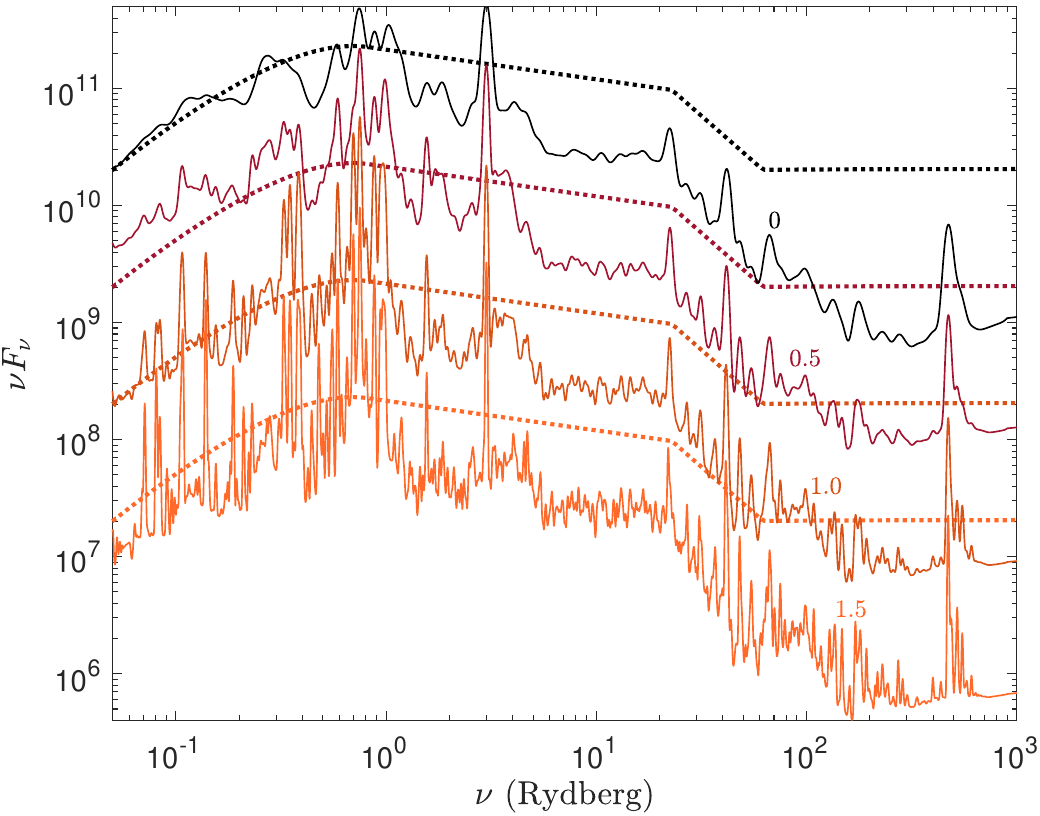}
\caption{Inwardly emitted spectra through the illuminated surface of a BLR cloud. Different curves correspond to the emission of clouds at different distances, $r,$ from the ionizing source.\ The numbers correspond to log($r/r_\mathrm{in}$). Dotted lines show the ionizing SED used for the calculations, relative to which full coverage BLR emission is assumed.}
\label{BLRlayers}
\end{figure}

The equilibrium structure of the cloud is self-consistently calculated here using the {Cloudy \sc v17.02} photoionization code (\citealt{Ferland2017}) including all relevant bound-bound, bound-free, and free-free opacities, as well as electron scattering. A dust-free medium is assumed, as appropriate for much of the line-emitting BLR (\citealt{Netzer1993}). The cloud's structure depends little on the boundary conditions at the location of the illuminated surface so long as the gas pressure at the boundary satisfies, $P_{\rm gas}(r) \ll F(r)/c $, where $F(r)$ is the total flux and $c$ is the speed of light. In our calculations, we consider geometrically thin cloud configurations so that the thickness of the emitting layer, $\delta r$, satisfies $\delta r/r<0.1$ \citep{Baskin2014}. In this limit geometrical attenuation inside the cloud is negligible. In all models considered here, the illuminated surface of the cloud is at or near the Compton temperature, which is $\gtrsim 10^6$\,K for a standard spectral energy distribution (SED) of a type-I quasar (\citealt{Chelouche2019}; see our Fig. \ref{BLRlayers}). The gas composition is characterized by a metal-to-hydrogen ratio of 2.5 times the solar ratio (\citealt{Warner2003}). The potential dependence of the ionizing SED on BH mass and accretion rate is ignored. Upon the calculation of the cloud's radial structure, its emissivity through the illuminated surface (i.e., the inwardly emitted spectrum) is obtained from infrared to X-ray energies (Fig. \ref{BLRlayers}). We assume that sightlines toward the observer do not pass through BLR clouds, and hence the adopted BLR configuration is akin to the bowl geometry (\citealt{Goad2012}), which is viewed face-on and is consistent with recent findings for the BLR geometry in low-luminosity sources \citep{Williams2018}.

\begin{figure}[t]
\includegraphics[width=8.9cm]{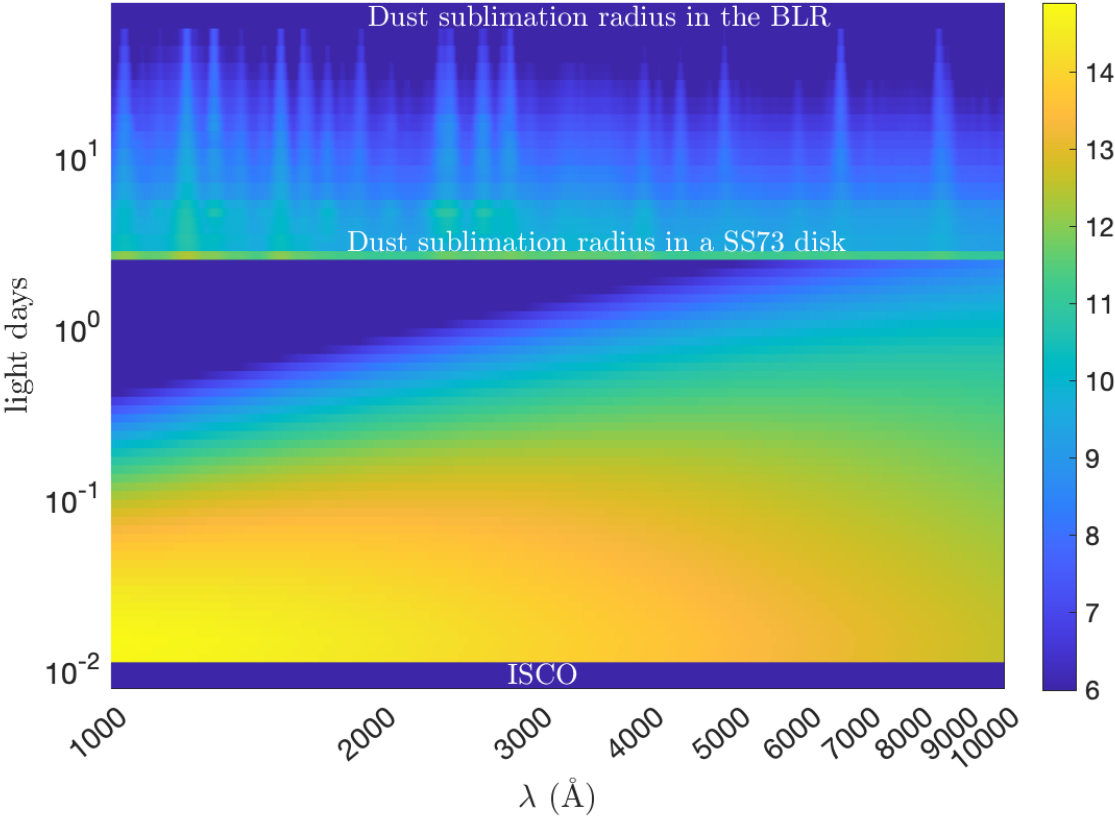}
\caption{Quasar SB as a function of the radial coordinate and rest wavelength for an $L_\mathrm{opt}=10^{43}\,\mathrm{erg~s^{-1}}$ source. Color-coding is proportional to the logarithm of the SB (see the colorbar using arbitrarily normalized units). Note the presence of the inner stable co-rotating orbit, the innermost stable circular orbit (ISCO), which is included here for a $\sim 10^7$ solar-mass BH.}
\label{BLRmodel}
\end{figure}

\begin{table*}
    \centering
        \tabcolsep=0.4cm
        \renewcommand{\arraystretch}{1.2}
        \caption{Emission line constraints for an $L_\mathrm{opt}=10^{43}\,\mathrm{erg~s^{-1}}$ source.}
        \vspace*{-3mm}
        \begin{tabular}{cccc|cccc} \hline \hline \vspace*{-3mm}\\
        & & & & \multicolumn{2}{c}{RPC model} & \multicolumn{2}{c}{mRPC model} \\
        Emission Line & EW & $R_\mathrm{BLR}$ & Baldwin &  EW & $R_\mathrm{BLR}$ &  EW &  $R_\mathrm{BLR}$\\ 
        & (\AA) & (light-days) & (index) & (\AA) & (light-days) & (\AA) & (light-days) \\ \hline \vspace*{-3mm} \\
        Pa$\beta$ & 40$\pm$20 & --- & 0.0 & 160 & 11 & 120 & 9\\
    H$\alpha$ & 390$\pm$120 & 15$\pm$6 & 0.0 & 250 & 13 & 180 & 9 \\
    H$\beta$ & 80$\pm$20 & 10$\pm$2 & 0.0 & 100 & 11 & 65 & 8 \\
    He\,II\,$\lambda 4686$ & $18\pm4$ & $2\pm1$ & -0.2 & 17 & 7 & 5 & 6 \\
    C\,III]+Si\,III]\,$\sim\lambda$1900  & 45$\pm$15 & 6$\pm$2 & -0.1 & 12 & 12 & 27 & 9 \\
    N\,III]\,$\lambda 1750$ & $8\pm3$ & --- & -0.1 & 32 & 10 & 24 & 7 \\
    He\,II\,$\lambda 1640$ & $22\pm5$ & $2\pm1$ & -0.2  & 22 & 7 & 20 & 6\\
    C\,IV\,$\lambda1548$ & $160\pm30$ & 3$\pm$1 & -0.2 & 190 & 9 & 100 & 7 \\
    Si\,IV$+$O\,IV]\,$\sim\lambda1400$ & 20$\pm$5 & 3$\pm$1 & -0.1 & 29 & 9 & 19 & 7 \\
    Ly$\alpha$ & 170$\pm$40 & 3$\pm$1 & -0.1 & 100 & 12 & 63 & 8 \\
    O\,VI\,$\lambda1035$ & 80$\pm$40 & --- & -0.2 & 120 & 9 & 100 & 7 \\
    O\,VIII\,$\lambda18.97$ & 0.05$\pm$ 0.02 & --- & --- & 0.1 & 6 & 0.06 & 6 \\
    Fe\,K$\alpha$\,$\lambda 1.93$ & 0.07$\pm$0.01 & --- & -0.2 & 0.03 & 6 & 0.01 & 6\\ \hline 
        \end{tabular}
\label{table}   
\end{table*}

\subsection{Model calibration}

The combined spectrum of many individual clouds depends on their spatial distribution. Here we assume that the gas covers a fraction of the quasar sky, which is quantified by a radially dependent covering fraction, $C(r)$, and that the quasar emits isotropically. Specifically, the cumulative covering fraction of the gas up to distance $r_\mathrm{in} \le r\le r_\mathrm{out}$ is given by
\begin{equation}
    C(r)=C_1 +C_2\int_{1}^{r/r_\mathrm{in}}\xi^\delta d\xi ,
\end{equation}
where $C_1$ and $ C_2$ are the parameters of the model. The power-law index, $\delta$ is rather loosely constrained by observations (certainly if $C_1>0$),   and we take $\delta=-1$ (\citealt{Baskin2014}), so that $C(r)=C_1+C_2\mathrm{ln}(r/r_\mathrm{in})$. Cases for which $C_1>0$ allow for the inclusion of an inner BLR "funnel," which is motivated by recent RM results (\citealt{Chelouche2019}), while $C_1=0$ is the more commonly employed model (\citealt{Baskin2014,Netzer2020}). Specifically, we define the RPC model with $C_1=0$ and the modified RPC (mRPC) model with $C_1=0.35$, which results in an inner funnel at the radius where a disk wind is launched (\citealt{Czerny2011,Chelouche2019}). This model is reminiscent of RPC models that are characterized by $\delta <-1$ values (\citealt{Netzer2020}). 

The outer BLR boundary, $r_\mathrm{out}$, is set by the sublimation radius for dust, such that $r_\mathrm{out}\simeq 10^3L_\mathrm{opt,45}^{1/2}$\,light-days (see \mbox{Fig. \ref{BLRmodel}})\footnote{Here we used the fact that the bolometric luminosity is $\simeq 8$ times the optical luminosity for the chosen SED.}. Realistically, a range of sublimation radii exists for different grain sizes and compositions, and yet a more accurate treatment of the problem is beyond the scope of the present work. Our neglect of BLR emission beyond $r_\mathrm{out}$ is justified by the large dust opacity at extreme-UV energies, which effectively competes with that of the gas over ionizing photons (\citealt{Netzer1993}). The suppression of gas emission is further enhanced by virtue of RPC models experiencing a much faster compression and hence being characterized by smaller emitting columns in the presence of dust, as these are inversely proportional to the effective absorption cross-section \citep{Baskin2014}. We neglect dust emission in our model, which is expected to contribute at wavelengths $\gtrsim 1\mu$m. Further, with $r_\mathrm{out}$ being typically significantly larger than the Einstein radius for our targets, the results are insensitive to the  particular value assumed. Here we take $r_\mathrm{out}/r_\mathrm{in}\sim 30$, which is based on arguments related to the effective launching of disk material to great heights above the AD surface beyond the dust sublimation radius in the disk (\citealt{Czerny2011}).

As our focus in this paper is on broadband emission from BLR gas, the particular kinematic model is immaterial, and simplified gas kinematics is assumed, whereby the velocity dispersion of the medium is set by the virial theorem, and Gaussian line profiles are assumed, except for Ly$\alpha$ where an approximate treatment of the extended wings is included (\citealt{Ferland2013}). The backward emitted spectrum emerging from the illuminated face of a cloud placed at several distances from the ionizing source shows emission lines, radiative recombination edges, and free-free emission covering a broad energy range (Fig. \ref{BLRlayers}). Notably, free-free emission combined with bound-free emission is substantial in the near-UV and optical bands, especially from gas that is located at $r\gtrsim r_\mathrm{in}$, where the substantial compression occurs, and gas densities exceeding $10^{12}\,\mathrm{cm^{-3}}$ are reached deep within the cloud layers. From such regions, Balmer line emission is collisionally suppressed (\citealt{Baskin2014,Chelouche2019}). Iron emission is substantial across the full spectral range (including the K$\alpha$ line at X-ray wavelengths) but is difficult to predict in the optical and UV bands, and should be cautiously interpreted.

The free parameters of the model are therefore $C_1,~C_2$, which are calibrated using ample reverberation and spectroscopic data pertaining to active galactic nuclei (AGN) with $L_\mathrm{opt,45}\gtrsim 0.01$, and are given in Table \ref{table}. Specifically, the equivalent width (EW) of rest-UV emission lines was taken from the \citet{Dietrich2002} sample for intermediate- to high-redshift sources of comparable luminosity. For the shortest wavelength emission lines (e.g., O\,{\sc VI}\,$\lambda 1035$) an extrapolation to low-luminosity sources was used, which is consistent with the data for local AGN (\citealt{Scott2005}). We note the lack of discernible redshift dependence in the EWs of prominent emission lines \citep{Dietrich2002} and take the scatter in the EW-redshift data as a measure of the EW uncertainty. The EW of prominent optical lines is affected by the host-galaxy contribution (\citealt{Stern2012,Shen2016}), and we adopt EW(H$\beta$)$=80\pm20$\AA\, as appropriate over a broad range of quasar luminosities (\citealt{Dietrich2002,Marziani2009,Shen2016}), and with an uncertainty that is consistent with the scatter seen in luminous sources for which the host galaxy contribution is small (\citealt{Marziani2009}). We take EW(H$\alpha$)$=$390\AA, which is based on an IR sample of luminous quasars (\citealt{Espey1989}) and consistent with recent composites for objects with a small host contribution to the flux (\citealt{Glikman2006,Stern2012,Selsing2016}). We scale the relative uncertainty in H$\alpha$ to that of H$\beta$. The EW of Pa$\beta$ is based on recent composites (\citealt{Glikman2006}) after correcting for a $\sim 50$\% host (and dust) contribution to the flux (\citealt{Selsing2016}) and conservatively estimating the uncertainty at the 50\% level. EW values for the X-ray lines were taken from X-ray grating observations of low-luminosity AGN (\citealt{Kaspi2002,Scott2005}). For more luminous sources, the Baldwin effect is observed, whereby the EWs of most emission lines decrease such that $\mathrm{EW} \propto L_\mathrm{UV}^\epsilon$, where $\epsilon\sim -0.2$ (\citealt{Baldwin1977,Dietrich2002,Page2004}).

\begin{table*}
        \tabcolsep=0.55cm
        \renewcommand{\arraystretch}{1.5}
        \caption{Half-light radius, $r_{1/2}$, estimates (at the 68\% confidence level) from the literature (scaled to  microlenses with a mean mass of 1 $M_\odot$).}
        \vspace*{-3mm}
        \begin{tabular}{c|cccc} \hline \hline \vspace*{-5mm}\\
        Object & $r_{1/2}$ & Rest wavelength& Einstein Radius &  Reference\\
        & (light-days) & (\AA)& (light-days) & \\ \hline \vspace*{-5.5mm} \\
        \multirow{5}{*}{SDSS 0924+0219} & $<13.8$ & 1400 & \multirow{5}{*}{22.3} & \citet{Floyd2009} \\
         & $4.7_{-2.4}^{+3.5}$ & 1400 & & \citet{Rojas2020} \\
         & $21.7_{-9.9^a}$ & 1736 & & \citet{Jimenez2012}\\
         & $4.9_{-3.3}^{+5.3}$ & 2770 & & \citet{MacLeod2015} \\ 
         & $1.6_{-0.9}^{+1.9}$ & 2770 & & \citet{Morgan2010}\\ \hline
         \multirow{2}{*}{RXJ 1131-1231} & $2.2$ & 4000 & \multirow{2}{*}{17.8} & \citet{Dai2010} \\
         & $3.5_{-1.3}^{+2.0}$ & 4220 & & \citet{Morgan2010} \\ \hline
         SDSS 1138+0314 & $1.3_{-0.9}^{+4.2}$ & 2030 & 21.1$^b$ & \citet{Morgan2010} \\ \hline
         \multirow{4}{*}{HE 0435–1223} & $15.3_{-4.7}^{+5.9}$ & 1310 & \multirow{4}{*}{20.7} & \citet{Motta2017} \\
         & $9.5_{-5.1}^{+12.0}$ & 1736 & & \citet{Jimenez2012}\\ 
         & $14.1_{-2.2}^{+12.8}$ & 2447 & & \citet{Fian2018} \\ 
          & $8.6_{-7.0}^{+18.6}$ & 2600 & & \citet{Morgan2010} \\ \hline
         \multirow{3}{*}{WFI 2033-4723} & $11.8_{-2.4}^{+3.5}$ & 1310 & \multirow{3}{*}{16.7$^b$} & \citet{Motta2017}\\ 
         & $7.7_{-4.7}^{+11.5}$ & 1736 &  & \citet{Jimenez2012} \\
         & $11.7_{-6.4}^{+9.2}$ & 2481 & & \citet{Morgan2010} \\ \hline
         \multirow{2}{*}{PG 1115+080} & $4.9_{-2.6}^{+5.7}$ & 1736 & \multirow{2}{*}{25.5$^b$} & \citet{Jimenez2012} \\ 
         & $68.5_{-41.3}^{+68.1}$ & 2570 & & \citet{Morgan2010} \\ \hline
         \multirow{2}{*}{WFI 2026-4536} & $<3.5$ & 1026 & \multirow{2}{*}{15$^b$} & \citet{Bate2018} \\
         & $5.2_{-2.5}^{+6.2}$ & 2043 & & \citet{Cornachione2020c} \\ \hline
         \multirow{4}{*}{MG 0414+053} & $<1$ & 1026 & \multirow{4}{*}{14.4} &\citet{Vernardos2018}\\
         & $<2.9$ & 1026 & & \citet{Bate2018}\\
         & $<8.2$ & 1808 & & \citet{Bate2008} \\ 
         & $<11.9$ & 2214 & & \citet{Bate2007} \\ \hline
        \end{tabular}\\ 
 
        $^a$ This value is a lower limit rather than a true estimate.\\
        $^b$ These values were taken from \citet{Mosquera2011}.
\label{sizes}   
\end{table*}

\begin{figure}
\centering
\includegraphics[width=8.9cm]{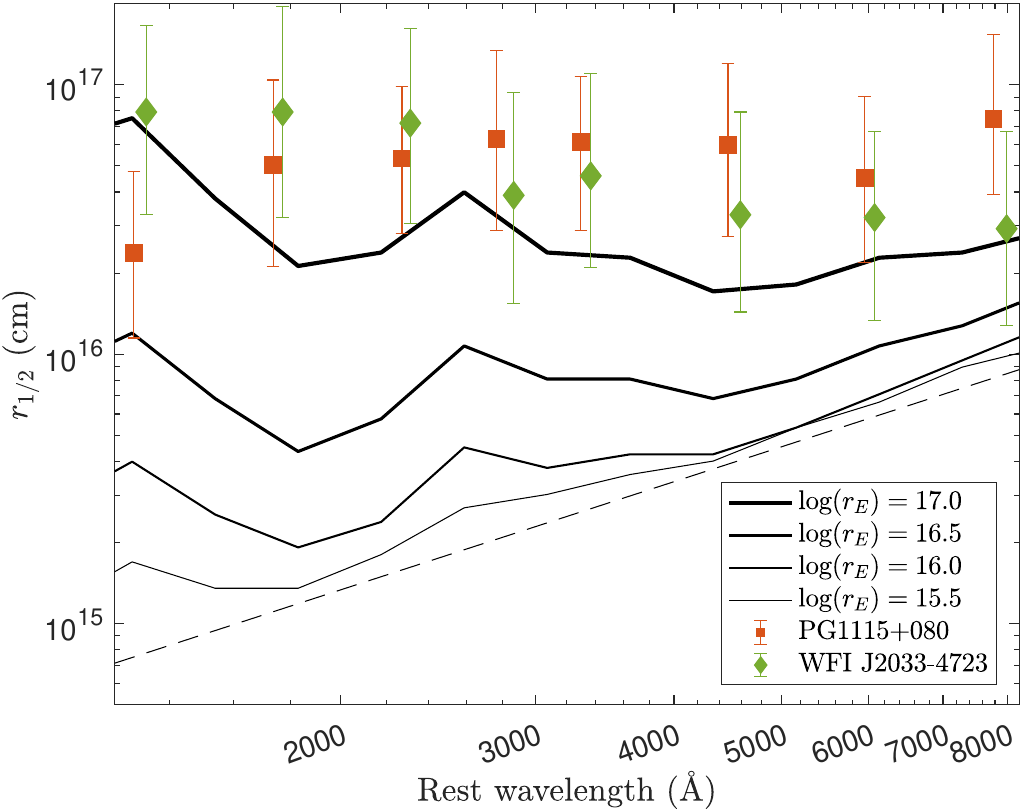}
\caption{Dependence of the deduced wavelength-dependent half-light radius, $r_{1/2}(\lambda)$, on the Einstein radius. The inferred $r_{1/2}(\lambda)$ from microlensing simulations are shown as black curves for different values of the Einstein radius, $r_E$ (see the legend). Overlaid are the microlensed-based size estimates for two $L_\mathrm{opt,45}\simeq 1$ quasars as well as the predicted SS73 wavelength-dependent disk size (dashed line).}
\label{R_E_bias}
\end{figure}

\begin{figure*}[t]
\centering
\includegraphics[width=18.3cm]{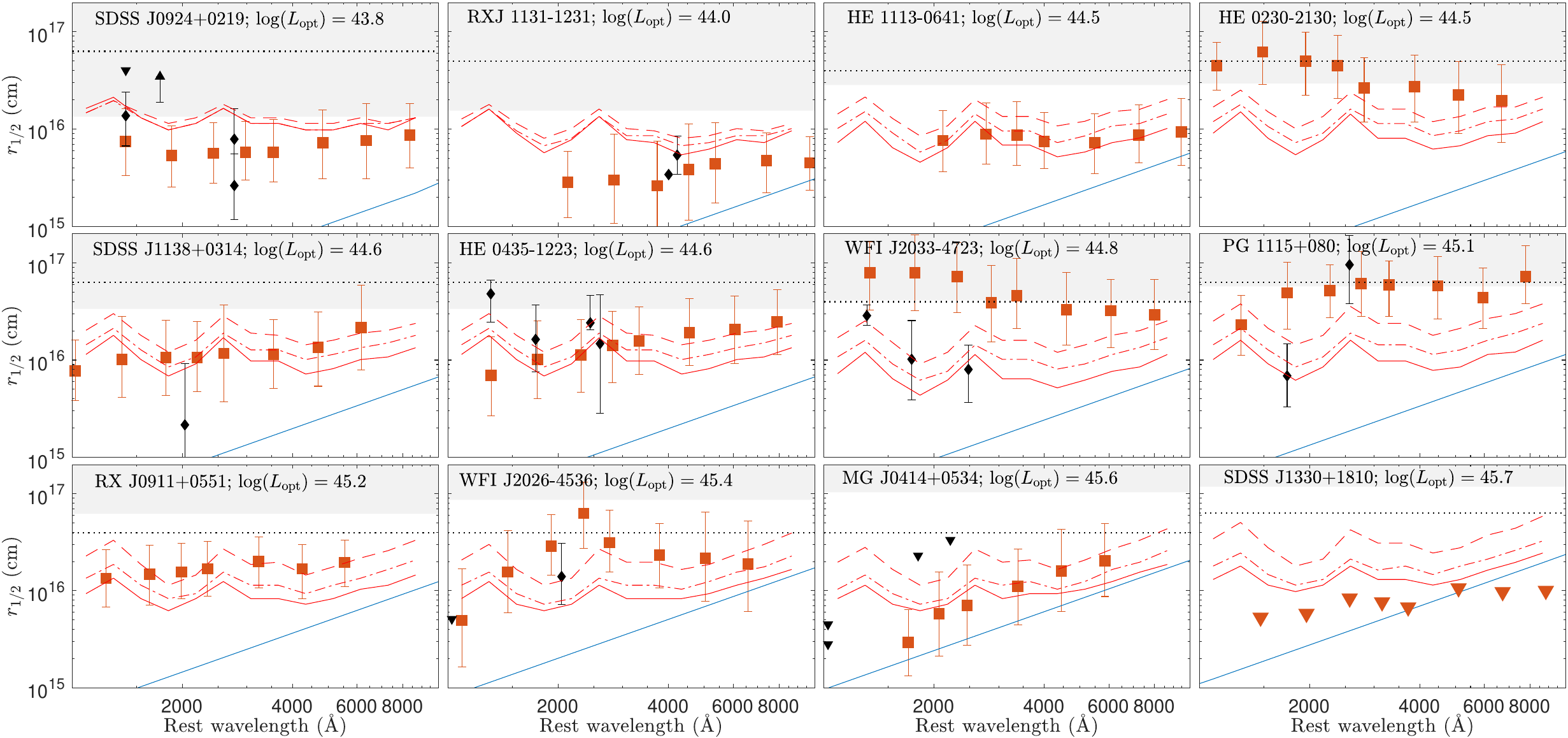}
\caption{Microlensing-based size estimates versus theoretical predictions for 12 strongly lensed quasars. The expected half-light radius for an SS73 disk for each source is shown as a blue line. The red curves show model predictions that include all relevant bound-bound, bound-free, and free-free emission processes and are convolved with a typical broadband throughput kernel to mimic broadband data. Three models are shown: a RPC model (\citealt{Baskin2014}) that accounts for the Baldwin relation (solid line), a similar model but with the addition of an inner BLR funnel (\citealt{Chelouche2019}; dotted-dashed curve), and a similar model to the latter but ignoring the effect of the Baldwin relation and setting the BLR properties to those given in Table \ref{table}, which characterize low-luminosity sources (dashed curve). Tiles are sorted by increasing object luminosity from left to right and from top to bottom. Microlensing data are shown with respective error bars (68\% confidence intervals) as red squares (\citealt{Blackburne2011}) and as black points (see Table \ref{sizes}). Upper and lower limits are denoted by downward and upward pointing triangles, respectively. The Einstein radius for each lens is denoted by the horizontal black dotted line. Shaded regions mark the scale range subtended by the BLR ($r_\mathrm{out}$ exceeds the axis limits in all cases).}
\label{tile}
\end{figure*}

A further set of constraints on our models involves the characteristic size scale, $R_\mathrm{BLR}$, of the region that gives rise to the emission of a particular line (see Table \ref{table}). Thus, different emission lines are characterized by different $R_\mathrm{BLR}$ values with that of H$\beta$ often being taken as a representative size for the entire BLR (\citealt{Bentz2013}). Here, reverberation sizes were collected from the literature (\citealt{Bentz2013}) and rescaled by $L_\mathrm{opt}^{1/2}$ to correspond to a $L_{\mathrm{opt},45}=0.01$ source. Specifically, $R_\mathrm{BLR}(H\beta)=10$\,light-days, and $R_\mathrm{BLR}(H\alpha)=15$\,light-days were used (\citealt{Bentz2010,Bentz2013}). The size of the CIV\,$\lambda 1548$ emission region was taken to be identical to that of Ly$\alpha$ and SiIV\,$\lambda 1403$, and four times smaller than the H$\beta$'s (\citealt{DeRosa2015,Lira2018}). The CIII] $\lambda 1900$ emission line region was taken to be twice that of the CIV line (\citealt{Lira2018}). Sizes for HeII lines were rescaled from recent reverberation studies  (\citealt{Bentz2010,DeRosa2015}). Constraints for the MgII line are unreliable on accounts of blending with the iron blends, and hence not included in the study. We refrain from using the recent GRAVITY data for the size of the Paschen-series emitting region since a proper comparison between reverberation data and interferometric data is yet to be performed. Model predictions for the emissivity-weighted radius are then compared to the measured $R_\mathrm{BLR}$ values for the emission lines listed in Table \ref{table}, as are the EW predictions. An optimal solution, in the $\chi^2$ sense, is sought in the phase space, whose predictions are given in Table \ref{table}.

Final model parameterization yields $C(r_\mathrm{out})=0.35$ for the RPC model, and $C(r_\mathrm{out})=0.7$ for the mRPC model. The main difference between the models is the larger contribution to the UV-optical continuum from the BLR in the mRPC case, and therefore to slightly lower EW values of prominent lines. The SB of the best-fit mRPC model is shown in Fig. \ref{BLRmodel} as a function of the radial coordinate and rest wavelength for a $L_\mathrm{opt,45} = 1$ source. 

Scaling the model to luminous quasars is done here by assuming that the system size scales with $L^{1/2}$ (\citealt{Bentz2013,Fian2023}, and below). Unless otherwise stated, we assume that the covering fractions, $C_1$ and $~C_2$, are proportional to $L^\epsilon$ with $\epsilon=-0.2$, which is motivated by the observed Baldwin effect (\citealt{Baldwin1978}; see our Table \ref{table}). Thus, in our model, the relative contribution of both line emission and non-disk continuum emission to the total flux is lower in more luminous quasars since the total covering fraction of the BLR over the quasar sky is smaller\footnote{The EW of the Balmer lines is less affected since the dominant free-free emission and bound-free emission at optical wavelengths is similarly suppressed.}. This is qualitatively consistent with the decreasing fraction of obscured to non-obscured sources with source luminosity (\citealt{Merloni2014}) that may be akin to the receding torus model (\citealt{Lawrence1991}). Specifically, the mRPC model with its inner funnel predicts a covering fraction of $\sim 0.3$ for our fiducial quasar, which is comparable to the observed fraction of X-ray and optically obscured sources among the quasar population (\citealt{Merloni2014}). We note, however, that we do not attempt to provide a physical explanation for the Baldwin effect, which is likely driven by additional factors relating, for example, to the shape of the ionizing spectrum and the gas metallicity (\citealt{Korista1998}).  

Assuming a face-on azimuthally symmetric source configuration, we convolve the wavelength-dependent SB for our fiducial, $L_\mathrm{opt,45}=1$ quasar with a magnification map, and estimate the source size as per the approach defined in Section \ref{mlm}. The recovered wavelength-dependent $r_{1/2}$ is shown in Fig. \ref{R_E_bias} after degrading the spectral resolution to qualitatively match broadband data. Predictions were calculated for several microlensing maps whose Einstein radii were set to an observationally motivated range of values ($\mathrm{log}(r_E)\in[15.5,17.0]$). Significant deviations from the input AD's half-light radius are apparent in all cases. Due to the different scales characterizing the quasar's SB profile, which involve the AD and the BLR, estimations of the wavelength-dependent half-light radius that are based on SS73 models do not simply scale with the Einstein radius. Specifically, the larger the Einstein radius is compared to the relevant physical scales in the SB map, the larger is the overestimation factor for the disk size. This is particularly noticeable at short wavelengths, which tends to flatten the size-wavelength relation, as is observed in quasars with comparable luminosity (see Fig. \ref{R_E_bias}; \citealt{Blackburne2011}). As a corollary, good understanding of quasar structure may be useful for assessing the mean mass of stars in the lens. 

\subsection{Comparison to the Blackburne et al. (2011) sample}

Microlensing-based size estimates for the continuum-emitting region in quasars were collected from the literature, where $\sim$90 wavelength-dependent size measurements for 12 strongly lensed quasars were augmented by $\sim$20 additional measurements for the same sources from various sources (Table \ref{sizes}). All measurements were scaled to an Einstein radius of a solar-mass star to allow the data sets to be compared. Specifically, the main sample analyzed here is based on the data provided in \citet{Blackburne2011}. Table \ref{sizes} presents additional size measurements for eight systems in that sample, along with their corresponding rest wavelength. Conversions have been made when necessary to translate all reported scales to SS73-disk half-light radii (as opposed to, e.g., Gaussian or uniform disk), and these are shown in Fig. \ref{tile}. We find that the size estimates drawn from different studies using different methods are consistent in most cases (compare the red and black points in Fig. \ref{tile}).

We next compare detailed model predictions to published microlensing estimations for the half-light radius of the continuum-emitting region for the above sources. Models are rescaled according to the magnification corrected luminosity for individual lensed quasars (Eq. \ref{r1/2}), and are then convolved with a magnification map whose Einstein radius matches the one provided for each source in the sample, which are all in agreement within 0.2\,dex. Model predictions for the recovered wavelength-dependent sizes for each source are shown in Fig. \ref{tile} (red curves), where the salient features are qualitatively reproduces, namely, a flatter size-wavelength dependence than assumed by pure AD theory (blue line), and implied AD sizes that are considerably larger than predicted by theory, especially at short wavelengths due to the BLR contribution to the bands. Predictions for three model variants for the BLR are provided, which differ by their treatment of the Baldwin effect ($\epsilon=-0.2$ vs. $\epsilon=0$) and/or the presence of an inner BLR funnel as motivated by recent reverberation results (RPC model with $C_1=0$ vs. an mRPC model with $C_1=0.35$). 

All models are qualitatively consistent with the half-light source sizes as inferred from microlensing data (red and black data points), although the model that does not account for the Baldwin effect and assumes as strong emission lines as in low-luminosity sources appears to lie above the short-wavelength data for the two brightest objects in the sample. At long wavelengths, microlensing size estimates are less biased and tend to converge to the predicted SS73 disk sizes based on the sources' luminosity in the more luminous objects. We note that in the case of SDSS J1330, the theoretical predictions lie significantly above the upper limits on the disk size, perhaps due to an overestimation of the quasar luminosity, or due to a less probable microlensing configuration, or a combination thereof (this source is unique in the sample of \citet{Blackburne2011} as it does not have corresponding X-ray data to further constrain the microlensing model). 

\section{Conclusions}\label{summary}

The comparison of our simulations to available data in the literature points to a possible solution to the oversized AD conundrum, which is related to the interpretation of the observed signals rather than to disk physics. Current AD and BLR models, limited and uncertain as they may be, are in qualitative agreement with microlensing-deduced continuum emission-region sizes. While additional causes for size overestimation exist, such as the effect of the BLR reverberation signal (see \citealt{Paic2022}) on forward modeling schemes (\citealt{Kochanek2004}), we show that the mere contribution of the BLR to the continuum signal is able to largely account for the implied disk overestimation factors as well as to give rise to the flat size-wavelength relations observed in a sample of 12 lenses. From this, combined with recent RM studies of low-luminosity sources (\citealt{Chelouche2019}), a coherent picture emerges wherein the BLR contribution to the signal must be accounted for before any constraints can be placed on  AD physics. This conclusion seems to apply whether the total nuclear emission from the quasar is being analyzed, as is the case with microlensing, or only its variable component, as is the case for RM studies.

Our findings appear to significantly weaken the tension between AD theory and observations. They suggest that the steep disk temperature profiles found by microlensing studies are erroneous as the data are largely affected by the BLR, which does not emit as an aggregate of black bodies and hence does not obey a temperature-wavelength relation. The BLR models considered here further suggest that microlensing is a promising tool for constraining the BLR physics, and in particular for assessing the elusive contribution of the BLR to continuum emission from quasars. Microlensing may therefore shed light on the long-sought origin of the Baldwin effect (\citealt{Baldwin1977,Baldwin1978}) by probing the dependence of the diffuse BLR continuum emission on the source luminosity and emission-line phenomenology, and revealing the mechanisms underlying the quasar eigenvector phenomenology (\citealt{Sulentic2000}). Shedding light on the above is crucial for revealing the inner workings of quasars, with implications for their usage as proxies of the BH census and, perhaps, as standard cosmological rulers. Our results indicate that microlensing may already provide useful constraints on AD physics in sources whose BLR emission is weak and its extent much larger than the typical Einstein radius, such as in the most luminous sources (see, e.g., MG\,J0414 in Fig. \ref{tile}). In particular, microlensing could shed light on the role of metallicity in setting the inner disk properties \citep{Jiang2016} as well as on the effects of both outflows from the disk atmosphere and the radiative transfer in the disk atmosphere on the disk phenomenology \citep{Laor2014,Hall2018}. Microlensing may also help elucidate the recently proposed connections between the AD, the BLR, and the torus in quasars \citep{Czerny2011,Czerny2016,Baskin2018,Dorodnitsyn2021}. Conversely, with quasar physics better established, it may be possible to independently estimate the mean mass of microlenses in the lensing galaxy from the shape and amplitude of the size-wavelength relation for quasars. This has implications for probing stellar populations in the outskirts of the lensing objects. With upcoming all-sky surveys, mapping quasar interiors using microlensing is expected to become widespread, and data may be able to tap into higher moments of the magnification distribution, potentially alleviating some of the biases and uncertainties discussed here.

\begin{acknowledgements}
We thank E. Mediavilla for generously providing the IPM code that enabled us to compute the microlensing magnification maps. We thank an anonymous referee for helpful suggestions that improved the conclusions section. This research was supported by the grant PID2020-118687GB-C32, financed by the Spanish Ministerio de Ciencia e Innovación. Research by D.C. and S.K. is financially supported by the German Science Foundation (DFG) grants HA3555-14/1, CH71-34-3, and by the Israeli Science Foundation grant no. 2398/19. Computations made use of high-performance computing facilities at the University of Haifa, which are partly supported by a grant from the Israeli Science Foundation (grant 2155/15).
\end{acknowledgements}

\bibliographystyle{aa}
\bibliography{bibliography}
\end{document}